# Relative Sensitivities and Correlation of Factors Introducing Uncertainty in Radiotherapy Dosimetry Audits

P. Krishnadas[†], S. A. Thomas, J. Goldring, N.A.S. Smith and M. Hussein,

*National Physical Laboratory,
London, TW11 0LW, United Kingdom*
[†]*E-mail: padmini.krishnadas@npl.co.uk*
*www.npl.co.uk*

Dosimetry audits are carried out to determine how well radiotherapy is delivered to the patient. It is also used to understand the uncertainty introduced into the measurement result when using different computational models. As measurement procedures are becoming increasingly complex with technological advancements, it is harder to establish sources of variability in measurements and understand if they stem from true differences in measurands or in the measurement pipelines themselves. The gamma index calculation is a widely accepted metric used for the comparison of measured and predicted doses in radiotherapy. However, various steps in the measurement pipeline can introduce variation in the measurement result. In this paper, we perform a sensitivity and correlation analysis to investigate the influence of various input factors (i.e. setting) in gamma index calculations on the uncertainty introduced in dosimetry audits. We identify a number of factors where standardization will improve measurements by reducing variability in outputs. Furthermore, we also compare gamma index metrics and similarities across audit sites.

## 1. Introduction

Measurement pipelines are becoming increasingly complex with technological advancements making it difficult to establish where exactly uncertainty in measurements stems from. Radiotherapy is a major modality for the treatment of cancer. The implementation of audits in radiotherapy pipelines is important to assure quality and safety in radiotherapy to patients[1]. By using tissue-mimicking material (or phantoms) in combination with passive or active detectors, it is possible to compare the measured radiation dose with the dose predicted by the hospital, providing a validation of how well the radiotherapy equipment and software have been implemented. Conducting these audits using the same measurement equipment in multiple centers allows for benchmarking and improving standards. One detector that is commonly used is called radiochromic film[2]. In an audit setting, measurements from the detector can be compared to the predicted dose generated by a radiotherapy Treatment Planning System (TPS). TPS is software used to simulate the radiation dose deposition in patients and to design the optimal radiotherapy approach. Gamma index[3,4] analysis is typically used to compare film and TPS dose, as it is a dimensionless metric which gives a quality indicator on the agreement between these distributions. However, there are many factors that affect the uncertainty of the measured film dose.

Sensitivity analysis[5] examines the extent of impact of individual factors, i.e. measurement setting, and the effect of interaction between the factors. It does so by identifying the factors introducing most variability to the measurement result or output. The results of sensitivity analyses are then used as a proxy for uncertainty introduced by the respective factors, i.e. measurement settings. Understanding the impact of different factors on the measurement result can help ensure the audit methodology's accuracy and improve radiotherapy quality. In this paper, we conduct a sensitivity analysis of Gamma index calculations made using the in-house Versatile Independent Gamma Analysis sOftware (VIGO) on dosimetry audit data[3] collected from Stereotactic Ablative Radiotherapy treatment (SABR) performed on phantom objects.

## 2. Methods

### 2.1. *Gamma Index Calculations and VIGO*

VIGO is NPL's in-house software developed in MATLAB[6] for the analysis of radiotherapy dosimetry audit data, producing 12 outputs. There are four Gamma Index Criteria (GIC) metrics, which are reported as both a median Gamma value and a Gamma Passing Rate (%), corresponding to eight of the 12 output. The other four are measurements for the mean dose difference (%); median dose difference (%); distance to agreement (DTA) (mm) and center of mass (CoM) distance (mm). Each of the GICs are computed at a specified dose difference (%) and distance difference (mm) criterion, these are; $GIC1$: 5 % / 2 mm, $GIC2$: 3 % / 2 mm, $GIC3$: 2 % / 2 mm and $GIC4$: 5 % / 1 mm. Figure 1 represents the steps involved in NPL's dosimetry audit highlighting the complexity in the workflow. The steps in blue represent where uncertainty can be introduced and propagated, which can have a significant impact on the output. The factors, levels and their combinations required for the sensitivity analyses were decided based on these blue steps, illustrated in Figure 1. It is interesting to note that the main factors found to introduce variability to measurement results were those associated with the Region of Interest (ROI) where the radiotherapy was delivered.

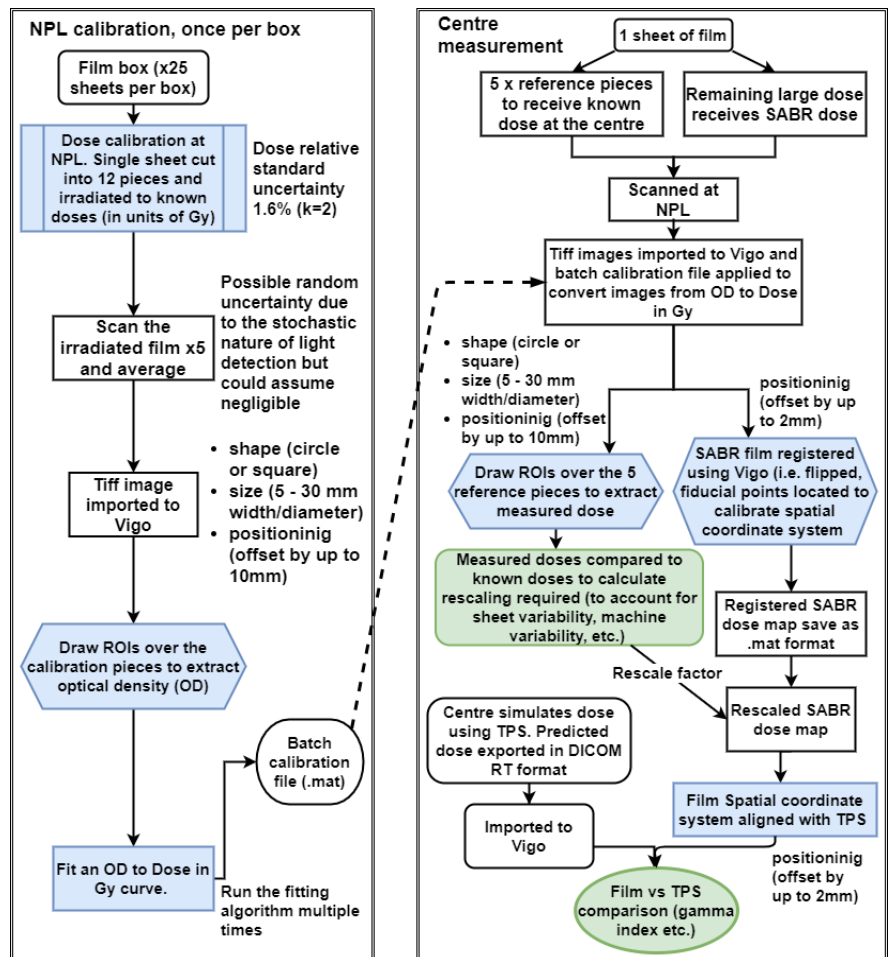

Figure 1.Steps involved in an NPL dosimetry audit, blue steps indicate the focus of our sensitivity analysis in this work. Green steps indicate steps where measurement results are collected. VIGO compares the delivered dosage to the reference dosage (available in the Treatment Planning System or TPS) to assess the quality of delivered radiation. The input is in the form of DICOM RT images (Digital and Imaging Communications in Medicine – Radiotherapy).

## 2.2. *Sensitivity Analysis*

Sensitivity analysis is a method used to identify and quantify the influential factors in an experimental design. Relative sensitivity[5,7] refers to the sensitivity of a measurement result to a factor in a measurement pipeline relative to its sensitivity to all other factors. We extend the concept of relative sensitivities by calculating the Type III sum of squares[7–9] to obtain a relative sensitivity value. The Type III sum of squares (also known as partial sum of squares) tests each term in the model against every other term in the model irrespective of order to calculate relative sensitivities for factors and their interactions over various levels. For further information about calculating relative sensitivities, we refer the readers to the following key texts[5,7–11].

## 2.3. *Experimental Design*

We analyze the output from VIGO based on various settings for the steps in blue boxes shown in Figure 1. This provides 12 outputs for each of the configurations of VIGO investigated for each center that has been audited. We calculate the relative sensitivities using a balanced design to provide insights on which factors are most influential in introducing uncertainties to measurement results, i.e. the VIGO metrics. We then compute the Pearson correlations between each of the metrics to understand the impact of different factors when using different GIC. This helped provide insights on which GICs produce measurements with similar characteristics. We also compute the correlation between factors themselves when using different GICs. This provided insights to which factors behave similarly in terms of introducing uncertainty.

## 3. Results

We performed a correlation analysis of the sensitivity analysis results as a proxy for uncertainties associated with factors. The sensitivity analysis results quantify the amount of variability introduced by each factor. Performing a correlation analysis on these factors after the sensitivity analysis allows us to understand which factors behave similarly in terms of introducing uncertainty. This in turn helps us understand if we can reduce uncertainty but controlling or standardizing specific factors. We can also ascertain if uncertainty can be reduced in several factors by standardizing a single factor; thus, reducing the variability introduced to the overall measurement result.

### 3.1. *Correlation between all centers for all metrics and measurements*

Across all centers and metrics, the results for the Gamma Passing Rate for each of the four GICs are strongly correlated with each other. Similarly, the mean Gamma metrics for different GICs are also closely correlated. The mean and median dose difference measurements are perfectly correlated with each other, however DTA and CoM are not correlated (Pearson's Coefficient = 0.65). Moreover, DTA and CoM difference measurement results are not correlated with any other measurement results, with the exception of DTA and Gamma Passing Rates for GIC1 and GIC4. Overall, this implies that measurement results from DTA and CoM are less likely to be influenced by the choice of GIC. These results are illustrated by the correlation matrix in Figure 2.

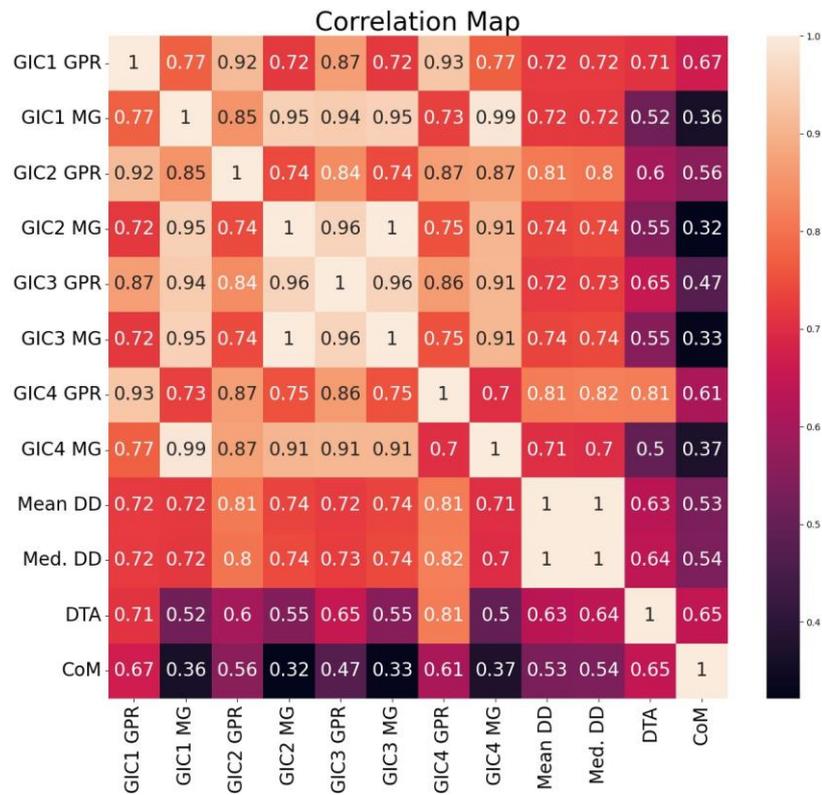

Figure 2.: Correlation matrix between different measurands for all centers. GIC=Gamma Index Criteria, GPR=Gamma Passing Rate, MG=median Gamma, DD=dose difference, DTA=distance to agreement, CoM=centre of mass distance.

### 3.2. *Correlation between centers for Gamma Passing Rate when using different Gamma Index Criteria*

Figure 3 is a matrix representing the correlation between centers for Gamma Passing Rates measurements made using GIC1. We observe that certain centers (such as c6 and c8) are closely related with a correlation score of 0.9 or above. We also observe some centers are poorly correlated with any of the other centers with correlation scores lower than 0.6. Specifically, we observe that c2 is poorly correlated with c1, c3, c4 and c7, while c5 is poorly correlated with c6, c8 and c9.

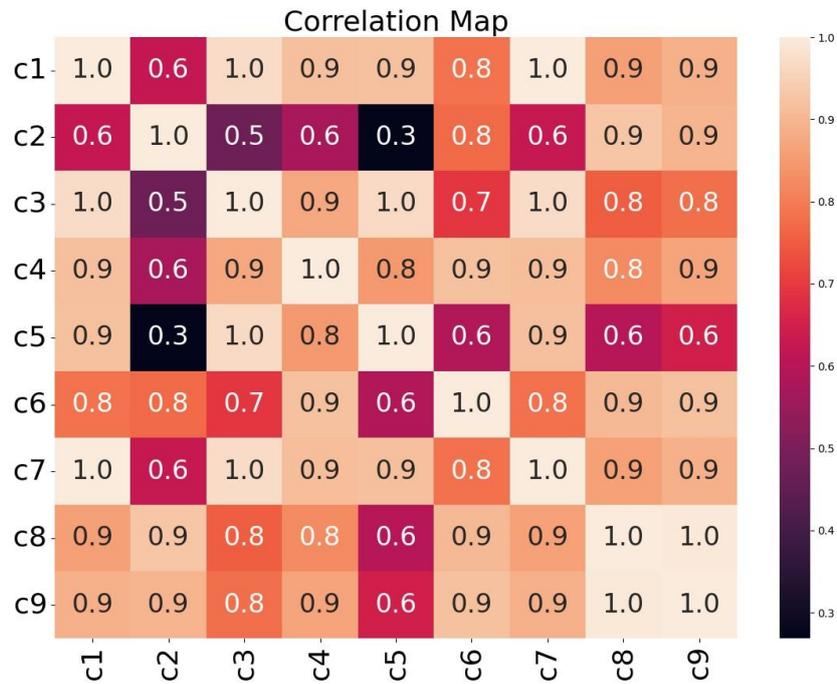

Figure 3.Correlation matrix between centers for Gamma Passing Rates (GPR) measurements when using GIC1. Centers are marked as c1, c2 etc.

We performed these correlation analyses for Gamma Passing Rates when using GICs 1, 2, 3 and 4. We found that c6 and c8; and c1 and c7 were strongly correlated and we can expect measurement results from these centers to have similar characteristics, whereas c2 and c5 were poorly correlated with most other centers when using all for GICs. This indicates that there are uncertainties within the measurement pipelines used at these centers, and they may benefit from further inspection in the interest of standardization of measurement results produced.

### 3.3. *Correlation between factors for Gamma Passing Rates using different Gamma Index Criteria*

Figure 4 is a correlation matrix representing the correlation between factors for Gamma Passing Rates made using GIC1. From Figure 4, we observe that the calibrated shape has a strong negative correlation with higher order interactions. This implies that if the uncertainty introduced to measurement results by the calibrated shape is small, uncertainty introduced by higher order interactions will be large. The calibrated size of region of interest was also observed to be negatively correlated with higher order interactions.

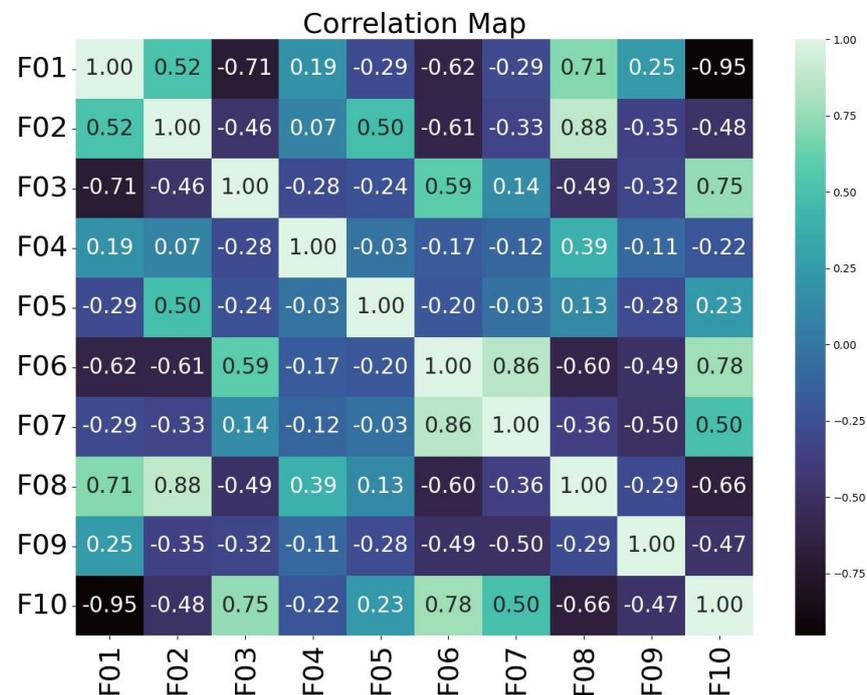

Figure 4. Correlation between factors for Gamma Passing Rates (GPR) made using GIC 1. Factors are indicated as: Calibrated shape = F01, Reference shape = F02, Calibrated offset dosage = F03, Calibrated offset on x axis = F04, calibrated offset on y axis = F05, reference offset on x axis = F06, reference offset on y axis = F07, Reference Size = F08, calibrated size = F09, higher order interactions = F10

We performed these correlation analyses for Gamma Passing Rates using GICs 1, 2, 3 and 4. We found that the calibrated shape and the calibrated size share a positive correlation while both the calibrated shape and size are negatively correlated with higher order interactions (nonlinear combinations of individual factors). Higher order interactions are positively correlated with the calibrated offset dosage, reference offset on the x and y axis.

## 4. Inference and Discussions

On interpreting the above analyses, we can better understand how the factors (measurement settings) in audits, centers conducting the measurements, and outputs investigated, can contribute to uncertainty introduced in radiotherapy audit measurements.

From the center correlation analysis, considering the trends displayed across the four GICs, we conclude that center 1 and center 7 display a strong positive correlation, as well as center 6 and center 8. This indicates that results from these centers may be produced from similar pipelines and may be similar in terms of introduced uncertainty in measurement results. However, center 2 and center 5 were poorly correlated with many other centers suggesting their results will diverge from regular trends. Measurement pipelines at these centers may benefit from further investigation to ensure standardization.

From the factor correlation analyses conducted and the trends displayed by the factors across the four GICs, we conclude that the calibrated shape and size of the regions of interest share a strong positive correlation. These two factors will have similar behavioral characteristics in terms of introducing uncertainty to measurement output. We also observe that the calibrated shape has a strong negative correlation with higher order interactions. This indicates that they do not behave similarly in terms of introducing uncertainty. Higher order interactions, however, were observed to be positively correlated with the calibrated offset dosage, reference offset on the x axis and the reference offset on the y axis. Therefore, standardizing these factors can help reduce uncertainty introduced not only through the factors directly, but also through the higher order interactions, and thus reduce over uncertainty associated with the measurement.

Determining the influence of each of the factors through sensitivity analysis and their correlation can help reduce the variability in measurement results. Understanding correlations between the factors highlights which factors are responsible for introducing variability to measurement results and influence on other factors. The correlation results help better ascertain which factors should be standardized (such as the shape and size of region of interest) as well as which factors should be controlled on priority to reduce the overall variability introduced to measurement results. Coupled with domain knowledge, we can use the correlation results to reduce the variability of factors that are difficult to control, e.g. higher order interactions, by standardizing factors that are correlated with that may be easier to control in measurements, e.g. calibration shape.

**Acknowledgements**

This work was funded by the UK Government's Department for Science, Innovation and Technology through the UK's National Measurement System programmes. The authors would also like to acknowledge Dr Xavier Loizeau and Dr Peter Harris from the National Physical Laboratory for their contributions to this paper.